\documentstyle[12pt,epsf,epsfig,wrapfig]{article}
\begin{document}
\phantom{xxxxxxx}

\sloppy
\vspace{5mm}
\begin{flushright}
FUB-HEP/96-16
\end{flushright}
\vspace{2.cm}
\begin{center}
{\large \bf
Left-right asymmetry and polarization for inclusive 
hyperon production processes\footnote{\normalsize Talk presented by C.~Boros 
at the 12th {\it International Symposium on High Energy 
Spin Physics}, SPIN96, Amsterdam, September 10-14, 1996.}
\footnote{\normalsize Supported in part by Deutsche Forschungsgemeinschaft 
(DFG:Me 470/7-2).}
 }
 
\vspace{5mm}
 C. Boros, Liang Zuo-tang and Meng Ta-chung
\\
\vspace{5mm}
{\small\it
Institut f\"ur theoretische Physik,\\ FU Berlin,
Arnimallee 14, 14195 Berlin Germany
\\ }
\end{center}

\begin{center}

\vspace{5mm}
\begin{minipage}{130 mm}
It is shown that the polarization 
in inclusive hyperon production and  the
left-right asymmetries in meson and hyperon production 
are closely related to one other. 
They can be
understood in terms of a 
picture in which the orbital motion of valence quarks
of the scattering hadrons play an important role.

\end{minipage}
\end{center}
\vfill

\newpage

Ever since the first hyperon-polarization experiments[1] people are
fascinated by the fact that hyperons produced in {\em unpolarized} 
hadron-nucleus collision processes are transversely {\em polarized}.   
Recently, the left-right asymmetries in 
inclusive $\Lambda$[2], $\pi^{\pm}$, $\pi^0$ and $\eta$ [3]  
production with polarized proton beams and
unpolarized proton targets has also  been measured.  
The characteristic features 
observed in both kinds of experiments 
are strikingly  similar. In fact:   
(1) both the polarization $P$ and the left-right asymmetries $A_N$ 
are significant, and {\em only} significant, in the fragmentation
regions   
of the colliding objects; (2) $P$ and $A_N$ depend on the flavor
quantum numbers of the produced particles (hyperons and mesons); 
(3) $P$ and $A_N$  depend 
on the flavor quantum numbers of the colliding objects. 

The above mentioned experimental facts suggest that both 
 phenomena are closely related to each other and  
that these phenomena have little to do with  typical hard 
scattering processes. Hence, it is  not surprising that    
the data contradict {\em perturbative} QCD where 
both $A_N$ and $P$ are expected to vanish. The experimental facts simply show 
that {\em soft,  
non-perturbative}  dynamics are needed in understanding them.  

In a recent paper[4], we proposed a 
model and pointed out the following:  
The existence of left-right
asymmetry for pion production can be understood on the basis of the
 theoretical assumptions and experimental facts mentioned above: 
(A) There exist a correlation between the polarization of the valence
quarks and their
transverse momentum distribution --- orbiting valence quarks. 
(B) Geometrical effects in hadron-hadron collisions play an important role
--- ``surface effect''. 
(C) Direct formation of hyperons
through fusion of valence (di)quarks of the projectile and
suitable sea-(di)quarks of the target are important 
in the fragmentation region
of the projectile. 
(D) u-quarks (d-quarks)  of a transverse polarized proton are
on average polarized in the same (opposite) direction as the proton 
(for details see Ref.4). 
In this talk I want to point out that also the left-right asymmetry for 
$\Lambda$-production can be understood and show that this picture 
naturally leads to hyperon polarizations in unpolarized production
processes, thus establishing a close relationship between these 
two spin phenomena. This talk is based on  the papers given in Ref.[5] 
written in collaboration with Liang Zuo-tang and Meng Ta-chung.

In order to understand the left-right asymmetry for hyperon 
polarization we recall that 
 there are the following three possibilities
for direct formations for $\Lambda$-production: 
(a) A $(u_vd_v)$-valence-diquark
from the projectile $P$ picks up
a $s_s$-sea-quark associated with the target
$T$ and forms a $\Lambda $:
$(u_vd_v)^P+s_s^T\to \Lambda $. 
(b) A $u_v$-valence-quark
from the projectile $P$ picks up a
$(d_ss_s)$-sea-diquark associated
with the target $T$ and forms a $\Lambda $:
$u_v^P+(d_ss_s)^T\to \Lambda $. 
(c) A $d_v$-valence-quark
from the projectile $P$ picks up a
$(u_ss_s)$-sea-diquark associated with
the target $T$ and forms a $\Lambda $:
$d_v^P+(u_ss_s)^T\to \Lambda $. 
We note that according to points (A), (B) and (D),
$\Lambda $ produced through
the direct formation process (b)
should have large probabilities to go left
and thus give positive contributions to $A_N$,
while those from $(c)$ contribute negatively to it.
For the direct formation process (a),
we note the following:
\begin{wrapfigure}{r}{8cm}
\epsfig{figure=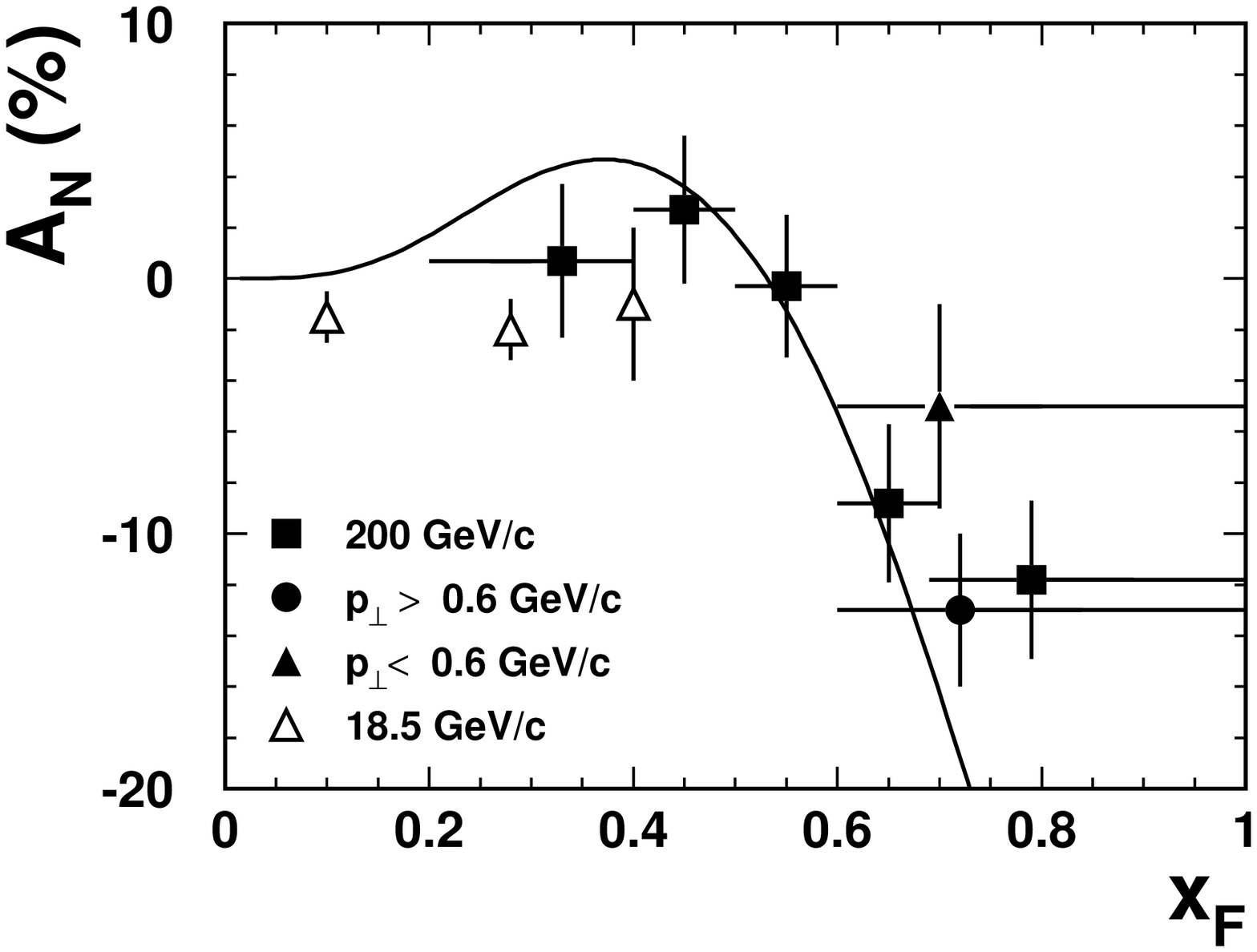,width=8cm}
{\small Figure 1: Left-right asymmetry $A_N$ for $p(\uparrow)+p(0)\to
\Lambda +X$ at $200$ GeV/c.
The data are from Ref. [6] and from Ref. [7]. 
}
\end{wrapfigure}
This direct formation process (a) should be
predominately associated with the production of
a meson directly formed through fusion of
the $u$ valence quark of the projectile
with a suitable anti-sea-quark of the target.
It follows from points (A),(B) and (D)
 that this meson should
have a large probability to
obtain an extra transverse momentum to the left.
Thus, according to momentum conservation,
the $\Lambda$ produced through (a) should
have a large probability to
obtain an extra transverse momentum to the right.
This implies that $(a)$ contributes negatively to $A_N$,
opposite to that of the associatively produced meson
($\pi ^+$ or $K^+$ or other) and 
 plays the dominating role
in the large $x_F$ ($x_F \ge 0.6$) region.
We therefore expect that
$A_N(x_F,\Lambda |s)$
is large  and negative  for large $x_F$ and for  
$x_F\sim 0.5$ it should be slightly positive.  
All these qualitative features are consistent
with the characteristics of the data[6,7]
(see Fig.1). 
\begin{wrapfigure}{r}{8cm}
\epsfig{figure=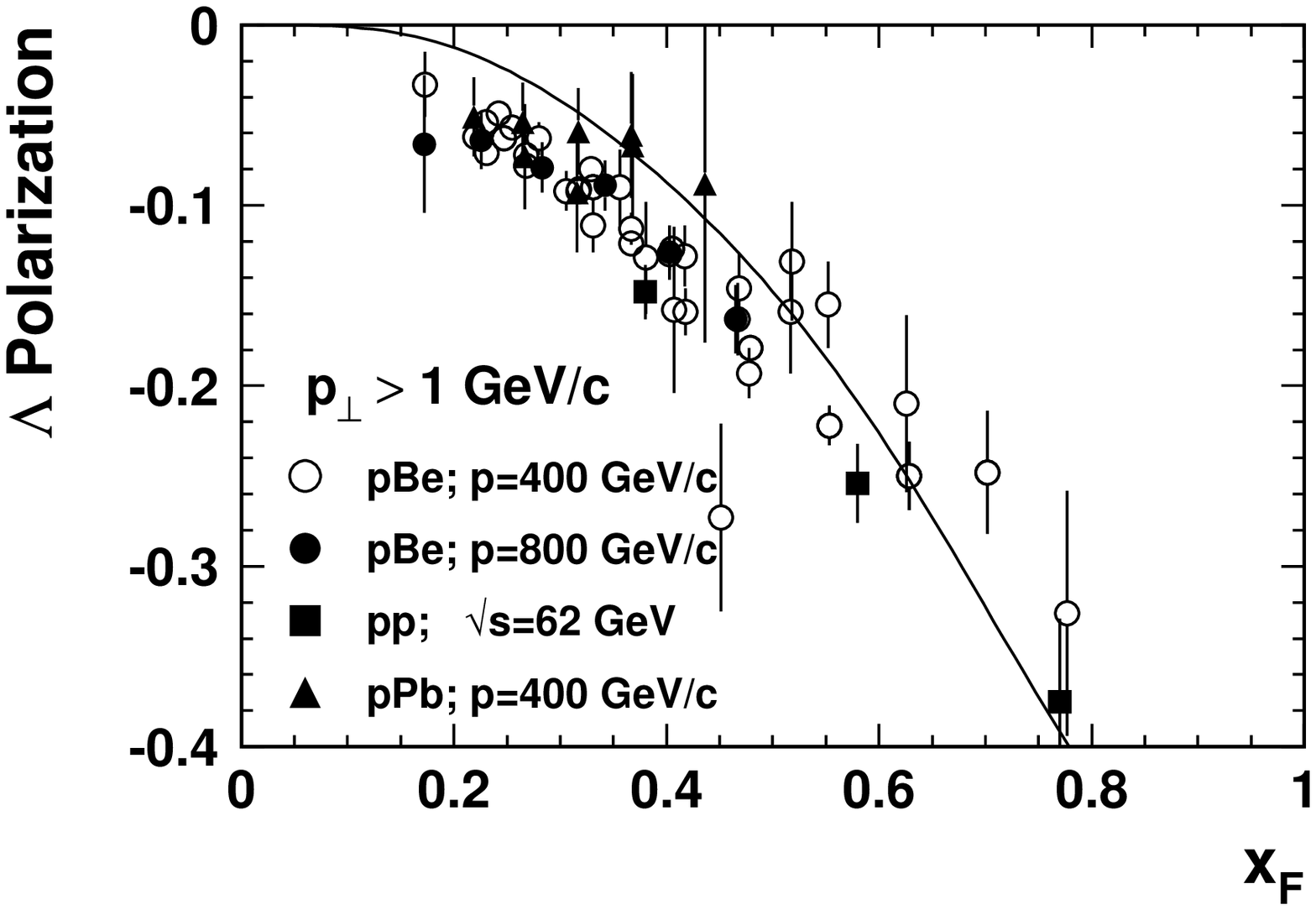,width=8cm}
{\small Figure 2: Polarization of $\Lambda$, $P_\Lambda$, as a function
of $x_F$.
Data are taken from [8].
The curve is the calculated result of the proposed model}
\end{wrapfigure}

Now we discuss the $\Lambda$-polarization and 
we first look at the $\Lambda$'s produced through (a).
This direct formation process
is mainly associated with
$(u_v^a)^P+\bar s_s^T\to K^+$.
Let us look at those
$\Lambda$'s which are going left.
In this case, according momentum conservation 
the associatively produced $K^+$
should have a large probability to go right.
This implies, according to (A), that $(u_v^a)^P$ has a large
probability to be downwards polarized. 
Note here, that by choosing a specific side (left or right)
with respect to the beam axis we select according to (A) and (B) a
certain polarization (up or down) although the projectile is not
polarized i.e. the prescription on which side we detect the
produced $K^+$ acts as a polarization filter. 
Since $K$ is a pseudo-scalar meson and
thus a spin-zero object,
$\bar s_s^T$ should be upwards polarized.
Hence, the corresponding $s_s^T$ should be downwards polarized,
provided that the sea quark-anti-quark pair is
not transversely polarized (see Ref. [5] for more details).  
Since the polarization of $\Lambda$ is
entirely determined by the $s$-quark, 
it follows that $\Lambda $ {\it should have a large
probability to be downwards polarized, i.e. $P_\Lambda <0$.} 
We note that this correlation remains unity even
when more pseudo-scalar mesons are created,
but may be destroyed if vector mesons
are associatively produced. 
In the following, we consider only the former case.
In this sense, what we obtain is the upper limit of
the expectation from the picture.
We also note that,
if $\Lambda$ is produced through (b) [or (c)] and is
going left, the valence quark $u_v$ [or $d_v$] should have a large
probability to be upwards polarized.
But this says nothing about
the polarization of $\Lambda$.
It means the $\Lambda$'s
formed through (b) and (c)
are not polarized.
This is why we expect to see
that $P_\Lambda$ is negative and significant only
in the large $x_F$ region (see Fig.2). 
According to the above picture $\Lambda$'s produced in diffractive
processes such us $pp\rightarrow (\Lambda K^+)p$ should have an even  larger
polarization since here momentum conservation
strictly implies that the $\Lambda$ and the $K^+$ have opposite
transverse momenta and the above mentioned correlation is also maximal.

\vspace*{1.cm}
{\Large \bf References}
\vspace*{0.5cm}

{\small\begin{description}

\item{[1]} A.~Lesnik {\it et al.,} Phys. Rev. Lett. {\bf 35}, 770 (1975);
G.~Bunce {\it et al.}, Phys. Rev. Lett. {\bf 36}, 1113, (1976).
\item{[2]}  See  K.~Heller talk given at this symposium 
and the papers cited therein.
\item{[3]} See  A.~Bravar talk given at this symposium 
and the papers cited therein.
\item{[4]} C.~Boros, Liang Zuo-tang and Meng Ta-chung,
               Phys. Rev. Lett. {\bf 70}, 1751 (1993);
\item{[5]} C.~Boros and Liang Zuo-tang, Phys. Rev. {\bf D53}, R2279
(1996) and 
``Orbiting valence quarks and polarization in inclusive production 
processes at high energies'' FU-Preprint FU-HEP/96-9. 
\item{[6]} A.~Bravar et al. Phys. Rev. Lett. {\bf 75}, 3073 (1995).
\item{[7]} A.~Lesnik et al.,
Phys. Rev. Lett. {\bf 35} (1975) 770;
B.~E.~Bonner et al. Phys. Rev. D {\bf 38} (1988) 729; 
\item{[8]} A.~M.~Smith {\it et al.}, Phys. Lett. {\bf 185B}, 209 (1987);  
B.~Lundberg {\it et al.}, Phys. Rev. {\bf D40}, 3557 (1989);  
E.~J.~Ramberg {\it et al.}, Phys. Lett. {\bf 338B}, 403 (1994). 
\end{description}}

\end{document}